\documentclass[twocolumn,prl,aps,showpacs]{revtex4}
\usepackage{amssymb}

\usepackage{epsfig}
\usepackage[english]{babel}
\usepackage{latexsym}
\usepackage{graphics}
\usepackage{subfigure}
\begin{document}
\title{Transition to Instability in a Kicked 
Bose-Einstein Condensate}
\author{Chuanwei Zhang$^{1,2}$, Jie Liu$^{1,3}$, Mark G. Raizen$^{1,2}$, and Qian Niu$^1$}
\begin{abstract}
A periodically kicked ring of a Bose-Einstein condensate is
considered as a nonlinear generalization of the
quantum kicked rotor.  For weak interactions between atoms, periodic motion (anti-resonance) 
becomes quasiperiodic (quantum beating) but remains stable. There exists a critical strength of 
interactions beyond which quasiperiodic motion becomes chaotic, resulting in an instability
of the condensate manifested by exponential growth in the number of noncondensed atoms.
Similar behavior is observed for dynamically localized states (essentially quasiperiodic motions), where stability
remains for weak interactions but is destroyed by strong interactions.

\end{abstract}
\address{$^1$Department of Physics, The University of Texas, Austin, Texas 78712-1081\\
$^2$Center for Nonlinear Dynamics, The University of Texas, Austin, Texas 78712-1081\\
$^3$Institute of Applied Physics and Computational Mathematics, P.O.Box 100088, Beijing, P. R. China}
\pacs{05.45.-a, 03.75.-b, 03.65.Bz, 42.50.Vk }
\maketitle
The classical kicked rotor is a textbook paradigm for dynamical chaos \cite{reichl}. 
The quantum kicked rotor has play an equally important role for the study of quantum chaos, for which a wide range of effects
have been predicted \cite{QKR} and observed in experiments \cite{raizen}. In recent years, the realization of Bose-Einstein 
condensation (BEC) of dilute gases \cite{BEC} has opened new opportunities for studying dynamical systems in the presence of many-body interactions.
A natural question to ask is how the physics of the quantum kicked rotor is modified by interactions \cite{NLSE}.

In the classical regime, chaotic motion leads to diffusive growth in the kinetic energy. In quantum dynamics, 
where chaos is no longer possible because of the linearity of the Schr\"{o}dinger equation, the motion becomes 
periodic (anti-resonance), quasiperiodic (dynamical localization), or resonant (quantum resonance) \cite{hog,anti-res}. 
In the mean field approximation, many-body interactions in BEC are represented by adding a nonlinear term 
in the Schr\"{o}dinger equation \cite{review}. This nonlinearity makes it possible to bring chaos back into the system,
leading to instability (in the sense of exponential sensitivity to initial conditions) 
of the condensate wave function \cite{instability}. The onset of instability of the condensate can 
cause rapid proliferation of thermal particles \cite{castin} that can be observed in experiments \cite{ketterle}. 
It is therefore important to understand the route to chaos with increasing interactions. This problem has recently been 
studied for the kicked BEC in a harmonic oscillator \cite{zoller}. 

In this Letter, we investigate the quantum dynamics of 
a BEC with repulsive interaction that is confined on a ring and kicked
periodically.  This system is a nonlinear generalization
of the quantum kicked rotor. From the point of view of dynamical theory, the kicked rotor is more generic than the kicked harmonic oscillator, 
because it is a typical low dimensional system that obeys 
the KAM theorem \cite{liu1}. It is very interesting
to understand how both quantum mechanics and mean field interaction
affect the dynamics of such a generic system.  We will focus most of our attention on the case of anti-resonance 
because it is the simplest periodic motion.  Here we find, with both analytic and numerical calculations, 
that weak interactions will make the motion quasiperiodic in the form of quantum beating. 
For strong interactions, quasiperiodic motions are destroyed and we observe a transition to instability of the
BEC that is also characterized by an exponential growth in the number of 
noncondensed atoms. Universal critical behavior for the transition is found. 
We have also studied nonlinear effect on dynamically localized states that may be regarded as quasiperiodic.  
Similar results are obtained in that localization remians for sufficiently weak interactions but become unstable 
beyond a critical strength of interactions. 

Consider condensed atoms confined in a toroidal trap of radius $R$
and thickness $r$, where $r \ll R$ so that lateral motion is
negligible and the system is essentially one-dimensional \cite{ring}. 
The dynamics of the BEC is described by the dimensionless
nonlinear Gross-Pitaveskii (GP) equation, 
\begin{eqnarray}
i\frac \partial {\partial t}\psi &=&\left( -\frac 12%
\frac{\partial ^2}{\partial \theta ^2}+g\left| \psi \right| ^2+K\cos\theta\delta_t(T)\right) \psi, \label{1}
\end{eqnarray}
where $g=8NaR/r^2$ is the scaled strength of nonlinear interaction,  $N$ is
the number of atoms, $a$ is the $s$-wave scattering length, $K$ is the kick
strength, $\delta_t(T)$ represents $\sum\limits_n\delta \left( t-nT\right)$, $T$ is the kick period, and $\theta $ denotes the azimuthal angle.
The length and the energy are measured in units $R$ and $\frac{\hbar ^2}{mR^2%
}$, respectively. The wavefunction $\psi \left( \theta ,t\right) $ has
the normalization $\int_0^{2\pi }\left| \psi \right| ^2d\theta =1$ and satisfies periodic boundary condition $\psi \left(
\theta ,t\right) =\psi \left( \theta +2\pi ,t\right) $. Experimentally, the ring-shape potential may be realized using two 2D
circular ``optical billiards" with the lateral dimension being confined by
two plane optical billiards \cite{billiard}. The 
$\delta $-kick may be realized by adding potential points along the ring
with an off-resonant laser \cite{raizen}.  The
interaction strength $g$ may be adjusted using a magnetic field-dependent Feshbach
resonance \cite{Feshbach}.

%%%%%%%%%%%%%%%%%%%%%%%%%%%%%%%%%%%%%%%%%%%%%%%%%%%%%
\begin{figure}[!t]
\begin{center}
\vspace*{-0.5cm}

\resizebox *{8cm}{7.5cm}{\includegraphics*{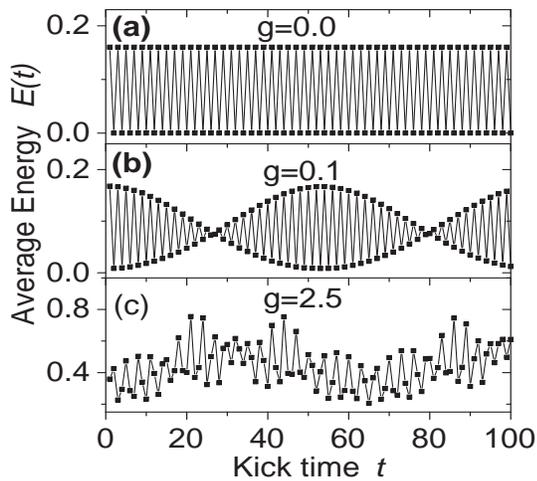}}
\end{center}
\vspace*{-1.0cm}
\caption{Plots of average energy $E(t)$ 
versus the number of kicks $%
t $ for three values of $g$. The kick strength $K=0.8$. }
\label{fig:gp}
\end{figure}
%%%%%%%%%%%%%%%%%%%%%%%%%%%%%%%%%%%%%%%%%%%%%%%%%%
The mean energy of each particle is 
$E(t)=\int_0^{2\pi }d\theta \psi ^{*}\left( -\frac 12%
\frac{\partial ^2}{\partial \theta ^2}+\frac 12g\left| \psi \right|
^2\right) \psi$. To determine the evolution of the energy, we
numerically integrate Eq.(1) over a time span of $100$ kicks, using a
split-operator method \cite{split}, with the initial wavefunction $\psi $
being the ground state $\psi \left( \theta ,0\right) =1/\sqrt{2\pi }$. The
kick period is chosen as $T=2\pi $ to match the condition for anti-resonance. After
each kick, the energy $E(t)$ is calculated and plotted as shown
in Fig.1. We see a remarkable difference among noninteraction
(Fig.1(a)), weak interaction (Fig.1(b)), and strong interaction (Fig.1(c)) 
cases. For noninteraction, the
energy $E(t)$ oscillates between two values (anti-resonance) and the period is $%
2T $; While in the weak interaction case, the amplitude of the
oscillation decreases gradually to zero and then revives, similar to the
phenomena of beating in classical waves. The values
of the \textit{oscillation} and \textit{beat} frequencies are obtained by
Fourier Transform and the
results are presented in Fig.2. It is shown that, besides the
appearance of the beat frequency, the interaction also modifies the
oscillation frequency; 
For stronger interaction  (Fig.1(c)), i.e. $g\ge1.96$, we find that
the energy's evolution demonstrates an irregular pattern, clearly 
indicating the collapse of the quasiperiodic motion and the 
occurrence of instability.
%%%%%%%%%%%%%%%%%%%%%%%%%%%%%%%%%%%%%%%%%%%%%%%%%%%%%
\begin{figure}[!b]
\vspace*{-0.5cm}

\begin{center}

\resizebox *{8cm}{4cm}{\includegraphics*{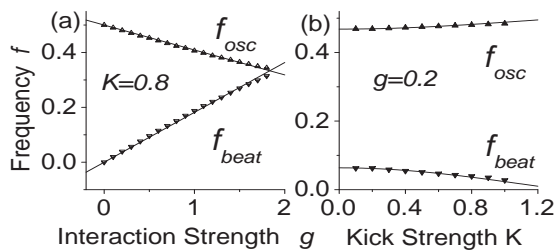}}
\end{center}
\vspace*{-1.0cm}
\caption{Plots of beat and oscillation frequencies versus the
interaction strength (a) and kick strength (b), where the scatters are the results from numerical
simulation using GP equation and lines from analytic expression Eq.(3).}
\label{fig:frequency}
\end{figure}
%%%%%%%%%%%%%%%%%%%%%%%%%%%%%%%%%%%%%%%%%%%%%%%%%%

%%

For the case of  weak interaction, the system is 
predominantly in the 
two energy levels. We can
map the system onto a spin model
with a two-mode approximation to the GP equation \cite{liu}.
By considering the conservation of parity we may write the wavefunction as 
$%
\psi =\frac 1{\sqrt{2\pi }}\left( a+\sqrt{2}b\cos \theta \right) $, where
the populations $a$ and $b$ at the ground and excited states satisfy the
normalization condition $\left| a\right| ^2+\left| b\right| ^2=1.$ The Hamiltonian in the spin representation  reads
\begin{eqnarray}
\hspace*{-0.5cm}
\mathcal{H} &=&-\frac {S_z}2+\frac g{2\pi}\left( S_x^2-\frac {S_z}4+\frac
{S_z^2}8\right) +\sqrt{2}KS_x\delta_t(T), \label{2}
\end{eqnarray}
where $S_z$ corresponds to the  population difference $\left| a\right| 
^2-\left| b\right| ^2$ 
and
arctan$(S_y/S_x)$ gives 
the  relative phase $\alpha =\arg (a)-\arg (b)$.
This
Hamiltonian is similar to a kicked top model \cite{kicktop},
but here the 
evolution between two kicks is more complicated.

With the spin model, we can readily study the dynamics of the system. 
For the case of noninteraction, the 
evolution between two consecutive kicks is
simply an angle $\pi$ rotation about the $z$
axis. 
The spin initially directing to north pole stays there 
for time duration $T$, then
the  first kick rotates  the spin 
by an angle $\sqrt{2}K$ about the $x$ axis. The 
following free evolution rotates the spin by an angle $\pi$ about
the $z$ axis.
Then, the second kick will drive the spin back to north pole through 
another rotation of $\sqrt{2}K$ about the $x$ axis.
With this the spin's motion is two kick period
recurrence and quantum anti-resonance occurs.

With interaction, the 
motion between 
two consecutive kicks is
approximately described by a rotation of $\pi+g(1+3S_z)/4$
about the $z$ axis. Compared with the noninteraction case, the mean field interaction leads to an additional
phase shift $g(1+3S_z)/4$. This phase shift results in a 
deviation of the spin from $S_x=0$ plane at time $2T^-$, i.e., moment just 
before the second kick. As a result, the second kick cannot drive the 
spin back to its initial position and quantum anti-resonance is absent. However, the phase 
shift will be accumulated in future evolution and 
the spin may reach $S_x=0$ plane
at a certain
time $mT^-$(beat period) when the total accumulated phase shift is $\pi/2$. 
Then the next kick will 
drive  spin back north pole by applying an
angle $\sqrt{2}K$ rotation about the $x$ axis.

The above picture is confirmed by our numerical solution of 
the spin Hamiltonian with
fourth-order Runge-Kutta method \cite{recipe}. 
In Fig.3 we see that, the relative phase
at the moment just before the even kicks increases almost linearly and
reaches $2\pi $ in  a beat period. 
The slope of the increment reads, 
$\gamma_{RP}=\left( \alpha \left( 4T^{-}\right) -
\alpha \left( 2T^{-}\right) \right) /2$, which 
can be deduced analytically. 
With this and through a lengthy deduction, we obtain  an analytic 
expression for the beat frequency to first order in $g$, 
\begin{eqnarray}
f_{beat} &\approx &\frac g{4\pi} \left( 1+3\cos\left(\sqrt2K\right)\right). \label{3}
\end{eqnarray}
This expression as well as the relation
between the oscillation frequency and the  beat frequency, $f_{osc}=\frac 12-\frac
12f_{beat}$, is in very good agreement with  
the numerical results as 
shown in Fig.2. Therefore the beating provides a method to measure interaction strength in an experiment.

%%%%%%%%%%%%%%%%%%%%%%%%%%%%%%%%%%%%%%%%%%%%%%%%%%%%%
\begin{figure}[!t]
\vspace*{-0.5cm}
\begin{center}
\resizebox *{8cm}{4cm}{\includegraphics*{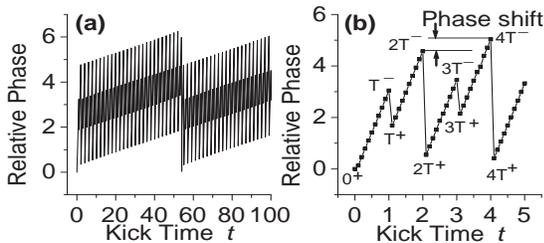}}
\end{center}
\vspace*{-0.7cm}

\caption{(a) Plots of relative phase versus the number of kicks $%
t $, where $g=0.1$, $K=0.8$. (b) Schematic plot of the phase shift. $nT^{-(+)}$ represents the moment just before (after) the $n$th kick.}
\label{fig:classical1}
\end{figure}
%%%%%%%%%%%%%%%%%%%%%%%%%%%%%%%%%%%%%%%%%%%%%%%%

Tuning the  interaction strength still larger means  
enhancing further the nonlinearity of the system. 
From our general understanding of nonlinear 
systems,  we expect that 
the solution will
be driven towards chaos, 
in the sense of exponential sensitivity to initial condition
and random evolution in the temporal domain. 
The latter character  has been  clearly displayed by  the   
irregular pattern  of  the energy evolution in Fig.1(c). 
On the other hand,
the onset of instability (or chaotic motion)
of the condensate is accompanied with the rapid proliferation of thermal
particles.
Within
the formalism of Castin and Dum \cite
{castin}, 
the growth of the number of the noncondensed atom will be exponential, 
similar to the exponential divergence of nearby trajectories in phase 
space of classical system. The growth rate of the noncondensed atoms is
similar to the Lyapounov exponent, turning from zero to  nonzero as  
instability occurs.

In Castin and Dum's formalism,
the mean number of noncondensed atoms at zero 
temperature is  described by $\langle
\delta \hat{N}(t)\rangle =\sum_{k=1}^\infty \langle v_k(t)|v_k(t)\rangle $,
where $|v_k(t)\rangle $ are governed by 
\begin{equation}
i \frac d{dt}\left(\! 
\begin{array}{c}
u_k\\ 
v_k 
\end{array}
\!\right) =\left(\!
\begin{array}{cc}
H+gQ|\psi |^2Q &\! gQ\psi ^2Q^* \\ 
-gQ^*\psi^{*2}Q &\! -H-gQ^*|\psi |^2Q^*
\end{array}
\!\right) 
\!\left(\!
\begin{array}{c}
u_k  \\ 
v_k
\end{array}
\!\right),  \label{4}
\end{equation}
where $H=%
\frac{\hat{p}^2}{2}+g|\psi |^2-\mu,$ $\mu $ is the chemical potential, $%
\psi $ is the ground state of GP equation and the projection operators $Q$ are given by $Q=1-|\psi \rangle \langle \psi |.$

We numerically integrate Eq.(4) for the $u_k$, $v_k$ pairs over a 
time
span of 100 kicks, using a split operator method, parallel to numerical
integration of GP equation. The initial
conditions $|u_k(0)\rangle $, $|v_k(0)\rangle ,$ for initial ground state
wavefunction $\psi (\theta )=1/\sqrt{2\pi },$ are obtained by 
diagonalizing the linear operator in Eq.(4) \cite{castin2}. After each 
kick
the mean number of noncondensed atoms is calculated and plotted 
versus time in Fig.4(a).
We find that there exists a critical value for the interaction strength, i.e.,
$g_c=1.96$, above which, the mean number of noncondensed atoms increases
exponentially, indicating the instability of BEC. Below the critical point,
the mean number of noncondensed atoms increases polynomially.  
As the nonlinear
parameter crosses over the
critical point,             
the growth rate
turns from zero to nonzero,
following   a square-root 
law (inset in Fig.4(a)).
This scaling law may be   universal for Bogoliubov 
excitation as  confirmed by recent  experiments \cite{ketterle}. 
%%%%%%%%%%%%%%%%%%%%%%%%%%%%%%%%%%%%%%%%%%%%%%%%%%%%%
\begin{figure}[!t]
\vspace*{-0.5cm}
\begin{center}
\resizebox *{8cm}{4.8cm}{\includegraphics*{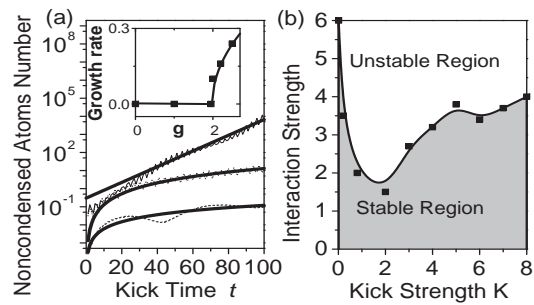}}
\end{center}
\vspace*{-1.0cm}

\caption{(a) Semilog plot of the mean number of noncondensed atoms versus the number of kicks $t$. 
The thicker lines are fitting functions. $K=0.8$, $g=0.1$ (dashed line, fitting
function $0.0003 t^{1.3} $), $g=1.5$ (dotted line, fitting function $%
0.0011t^2$), $g=2.0$ (dash dotted line, fitting function 0.32exp(0.1t)). 
The inset shows the interaction dependence of the growth rate. 
The scatters are from numerical simulation and the solid line is the fitting function $0.33{(g-1.96)}^{1/2}$.
(b) Phase diagram of the transition to instability.}
\label{fig:bog}
\end{figure}
%%%%%%%%%%%%%%%%%%%%%%%%%%%%%%%%%%%%%%%%%%%%%%%%%%

The critical value of the  interaction strength   depends on the 
kick strength.
For  very small kick strength, the critical interaction is expected to
be large, because the ground state of the ring-shape BEC with repulsive interaction
is dynamically stable \cite{biao}. For  large kick strength, 
to induce chaos,
the interaction strength must be large  enough to compete with the 
external kick potential.
So, in the parameter plane of $(g,k)$, the boundary
of instability shows a "U" type curve (Fig.4(b)).

Across the critical point, the  
density profiles of both condensed and noncondensed atoms
change dramatically.
In Fig.5, we plot the temporal evolution of the  density distributions of 
condensed atoms  as well 
as noncondensed atoms.
In the stable regime, the condensate density oscillates 
regularly with time and shows clear beating pattern (Fig.5(a)), whereas 
the  density of the  noncondensed atoms grows slowly 
and shows main peaks around  $\theta=\pm \pi$ and $0$, besides some
small oscillations (Fig.5(b)).
In the unstable regime, the temporal oscillation of the
condensate density is irregular (Fig.5(c)),
whereas the density of noncondensed atoms  grows explosively with the 
main concentration peaks at $\theta=\pm \pi/2$ where
the gradient density of the condensed part is maximum 
(Fig.5(d)).
Moreover,
our numerical explorations show  that 
the $\cos^{2}\theta$ mode (Fig5.(b)) dominates
the density distribution of the 
noncondensed atoms as the interaction strength  is less than 1.8. 
Thereafter, the
$\sin^{2}\theta$ mode grows while $\cos^{2}\theta$ mode
decays, and finally $\sin^{2}\theta$ mode 
become dominating in the
density distribution of noncondensed atoms above  the transition 
point (Fig5.(d)). 
Since the density distribution can be measured in experiment, this effect can be used to identify the transition to instability.
%%%%%%%%%%%%%%%%%%%%%%%%%%%%%%%%%%%%%%%%%%%%%%%%%%%%%
\begin{figure}[!t]
\vspace*{-0.5cm}

\begin{center}
\resizebox *{8cm}{6.14cm}{\includegraphics*{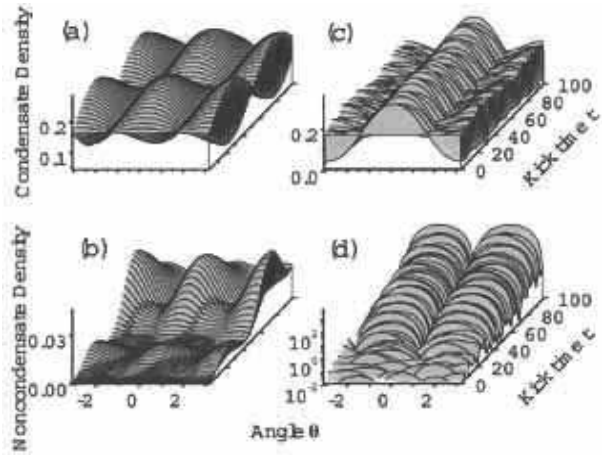}}
\end{center}
\vspace*{-0.5cm}

\caption{Plots of condensate and noncondensate densities, where $K=0.8$. (a,b) $g=0.1$; (c,d) $g=2.0$. }
\label{fig:wave}
\end{figure}
%%%%%%%%%%%%%%%%%%%%%%%%%%%%%%%%%%%%%%%%%%%%%%%%%%

%%%%%%%%%%%%%%%%%%%%%%%%%%%%%%%%%%%%%%%%%%%%%%%%%%%%%
\begin{figure}[!b]
\vspace*{-0.5cm}

\begin{center}
\resizebox *{8cm}{4cm}{\includegraphics*{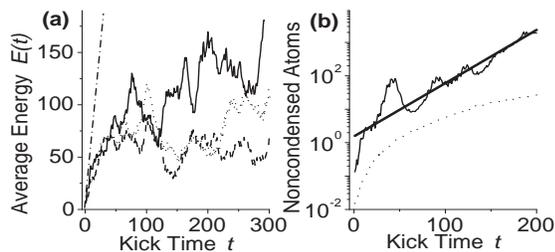}}
\end{center}
\vspace*{-0.8cm}

\caption{Nonlinear effects on dynamically localized states. $K=5$, $T=1$. (a) Plots of average energy $E(t)$ versus the number of kicks $t$, where
dash dotted line corresponds to the classical diffusion. $g=0$ (dash), $g=1$ (dot), $g=5$ (solid). 
(b) Semilog plot of the mean number of noncondensed atoms versus the number of kicks $t$. $g=1$ (dot),
$g=5$ (solid).}
\label{fig:dl}
\end{figure}
%%%%%%%%%%%%%%%%%%%%%%%%%%%%%%%%%%%%%%%%%%%%%%%%%%

Finally, although the above discussions have been focused on a periodic state 
of anti-resonance, the transition to instability due to strong interactions also follows a
 similar path for a dynamically localized state \cite{detail}.  The only difference is that we start out with a quasiperiodic
 rather than periodic motion in the absence of interaction.  This means that it will generally be easier to induce 
instability but still requires a finite strength of interaction.  In Fig.6, we show the nonlinear effect on a 
dynamically localized state at $K=5$ and $T=1$.     
For weak interactions ($g=1$) the motion is quasiperiodic with slow growth in the number of 
noncondensed atoms.  Strong interaction ($g=5$) destroys the quasiperiodic motion and leads to diffusive growth of energy, 
accompanied with exponential growth of noncondensed atoms that clearly indicates the instability of the BEC. 
Notice that the rate of growth in energy is much 
slower than the classical diffusion rate, which means that chaos brought back by 
interaction in this quantum system is still much weaker than pure classical chaos. 

We acknowledge the support from the NSF, and the R. A. Welch foundation, MGR
also acknowledges supports from the Sid W. Richardson Foundation.


\begin{references}
\bibitem{detail} More details will be provided in the future publication.
\end{references}
\end{document}